\documentclass{scrartcl}

\usepackage{amsmath}
\usepackage{color}
\usepackage{amsmath}
\usepackage{amssymb}
\usepackage{dsfont}
\usepackage{tikz}
\usepackage{subcaption}
\usepackage{float} 
\usepackage{fullpage}
\usepackage{overpic}
\usepackage{todonotes}
\def\be{\begin{equation}}
\def\ee{\end{equation}}
\def\bea{\begin{eqnarray}}
\def\eea{\end{eqnarray}}
\usepackage{breqn}

\begin{document}
\title{Phase field as a front propagation method for modeling grain growth in additive manufacturing}
\author{}
\author{
	\large
 Murali Uddagiri $^{1, *}$,  Pankaj Antala $^{2}$, and Ingo Steinbach $^{1}$\\
	[2mm] \small $^1$ ICAMS, Ruhr University Bochum, Universit\"{a}tsstr. 150, 44801 Bochum, Germany\\
               \small $^2$ Department of Materials Engineering, Katholieke Universiteit Leuven, Arenberg 44, Leuven, 3001, Belgium\\
	\vspace{-5mm}}
	
\maketitle

\begin{abstract}
\noindent
Abstract: A mesoscopic grain-envelope model applying a phase-field front-propagation method is developed to simulate grain growth under additive manufacturing process conditions. The envelope represents the outer surface of dendritic grains through a diffuse interface. While a modified heat-conduction model that incorporates moving heat sources and latent-heat release provides the evolution of local thermal field. Envelope propagation is determined from microscopic-solvability-based kinetic law. The model is validated through two- and three-dimensional simulations and subsequently applied to examine the influence of material and process parameters on microstructure evolution. The results demonstrate that the proposed mesoscopic model offers an efficient and predictive approach for modeling grain growth during multi-pass and multi-layer build-up in additive manufacturing.

\end{abstract}
Keywords: Phase field, Mesoscale, Graim growth, Texture, Additive Manufacturing. 

\section{Introduction}\label{intro}
Additive manufacturing (AM), particularly powder-bed fusion AM (PBF-AM), has gained significant attention for producing complex metallic components with tailored microstructures and mechanical properties. Microstructure development during PBF-AM is highly sensitive to process parameters and strongly influenced by rapid solidification driven by large thermal gradients and high interface velocities. These conditions promote non-equilibrium microstructural features, including epitaxial columnar grains with strong $\langle 001 \rangle$ fiber texture, which contribute to mechanical anisotropy {\textcolor{blue}{\cite{kok_anisotropy_2018}}}. Repeated thermal cycling during layer-by-layer deposition further induces phase transformations and intermetallic formation, complicating microstructure–property relationships. Process parameters such as beam power, scan speed, and layer thickness directly affect local solidification behavior, and variations in heat input across AM systems lead to significant differences in grain morphology even for the same alloy \cite{debroy_additive_2018, keller2017application}. These considerations highlight the need for computationally efficient and predictive microstructure models to support process optimization and material design in AM.

A number of numerical techniques including front-tracking, phase-field (PF), level-set (LS), lattice Boltzmann (LB), Potts kinetic Monte Carlo, cellular automata (CA), and Johnson–Mehl–Avrami–Kolmogorov-type approaches have been employed, each offering distinct advantages and limitations depending on computational cost and physical fidelity {\cite{gunasegaram_modelling_2021, tourret_comparing_2020}}.

Cellular automata (CA) models pioneered mesoscale simulations of competitive grain growth and texture evolution at modest computational cost \cite{gandin_analytical_1996, teferra_optimizing_2021}. Their coupling with finite element thermal solvers in cellular automata–finite element (CAFE) frameworks has been widely used for AM microstructure modeling, capturing epitaxial grain growth, competitive selection, and texture evolution \cite{zinovieva_three-dimensional_2018,teferra_optimizing_2021, rai_coupled_2016}. More recent CA implementations incorporate temperature-gradient tracking, grain competition, and probabilistic nucleation. For example, \cite{yang_sample3d_2023} combines a finite-difference thermal solver for electron-beam melting with CA to predict grain structures under varying process conditions. While efficient, CA-based methods rely on empirical rules for interface kinetics and cannot inherently capture curvature effects, solute redistribution, or thermodynamically consistent interface motion. Their predictive accuracy diminishes when solidification becomes strongly transient, such as under keyhole-mode melting or rapid inter-layer reheating.

Phase-field (PF) methods provide a thermodynamically rigorous description of diffuse interfaces and have become a leading approach for modeling alloy solidification \cite{steinbach2009phase,steinbach_phase-field_2013}. PF models enable quantitative simulation of dendritic growth, solute redistribution, and grain morphology in dilute and multicomponent alloys. Numerous studies have applied PF to AM-like rapid solidification, including melt-pool simulations in Ni-based superalloys, Ti–6Al–4V, steels, and Al alloys. In our previous work \cite{uddagiri_solidification_2023}, a multi-phase-field approach with constant heat extraction was used to investigate layer remelting, epitaxial growth, and solute-driven nucleation. However, AM microstructures are dictated by the tight coupling of heat transfer, melt flow, and solidification, motivating the development of concurrent multiscale frameworks. For example, thermal–lattice Boltzmann approaches have been used (in 2D) to jointly solve fluid flow and heat transfer \cite{korner_modeling_2020}, while \cite{chadwick_development_2021} employed an analytical Rosenthal solution to couple PF simulations with highly non-equilibrium solidification conditions.

Because the temperature field forms the foundation of any solidification model, AM simulations must resolve rapid, transient thermal evolution within small melt pools, which imposes severe computational constraints. Hierarchical strategies therefore remain common, in which macroscopic CFD or finite element models provide thermal gradients and cooling rates for mesoscale PF or CA simulations \cite{karayagiz_finite_2020, azizi_characterizing_2021, tang_phase_2022}. Although computationally feasible, these approaches require assumptions about interface kinetics and generally cannot resolve dendritic-scale curvature or solutal effects.

Despite significant progress, fully resolving solute diffusion and dendrite morphology over millimeter-scale AM melt pools remains computationally prohibitive. To address this, mesoscopic envelope models were proposed as an efficient alternative \cite{steinbach_three-dimensional_1999, steinbach_transient_2005}. These models represent a dendritic grain by a smooth envelope surface that tracks the outer boundary of active dendrite tips while incorporating physically meaningful growth kinetics. When integrated with PF front propagation, envelope models enable computationally tractable simulations of competitive grain growth and texture evolution under realistic AM thermal fields.

Building upon this concept, the present work develops a mesoscopic grain-envelope model coupled with a phase-field front-propagation method for PBF-AM. The framework incorporates a modified heat-conduction formulation including moving heat sources and latent heat, and employs a microscopic-solvability-based kinetic relation to compute interface velocities from local undercooling. The model is validated through two- and three-dimensional simulations and used to investigate the influence of material properties and temperature gradients on microstructural selection during AM.

\section{Model description}

\begin{figure}[h!]
    \centering
    \includegraphics[width = \textwidth,scale=0.4]{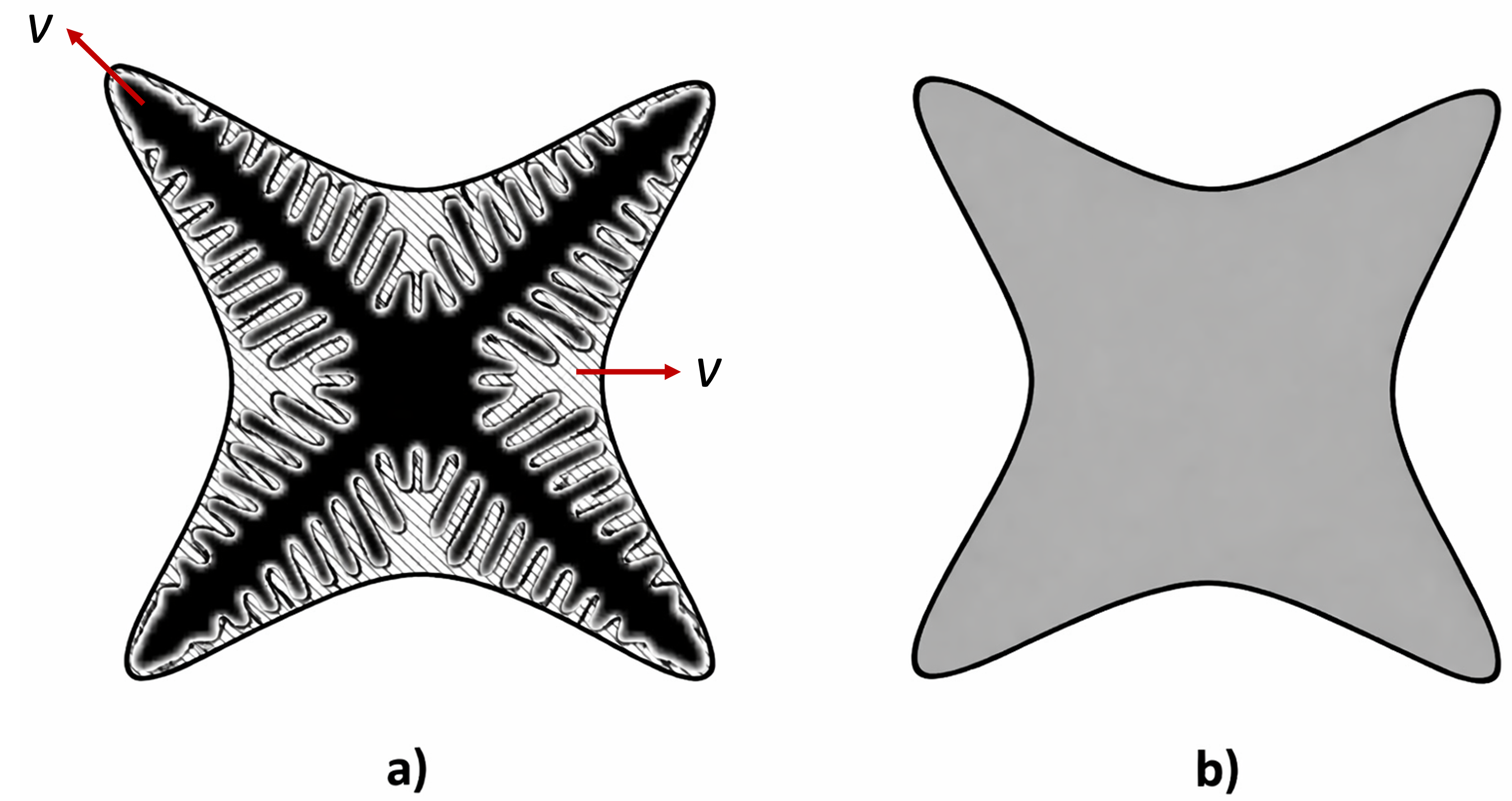}
    \caption{ Schematic illustration of a) Envelop around dendrite branches and b) Equivalent envelop surface, adapted from \cite{steinbach_three-dimensional_1999}}
    \label{fig:envelop}
\end{figure}

The phase-field based grain-envelope model, originally introduced by on of the authors and co-workers \cite{steinbach_three-dimensional_1999, steinbach_transient_2005} provides a scale-bridging approach for dendritic solidification. This approach connects the microscale (dendrite tip), grain-internal morphology, and the mesoscale interaction of multiple grains. As shown in the Fig.\ref{fig:envelop}, instead of resolving multiple dendrite arms within a grain, the entire multi-branched dendritic structure of a grain is represented by a smooth envelope surface that connects the active dendrite tips.  The region inside the envelope is assumed to be either fully solid or mushy, while the envelope surface evolves according to dendrite-tip growth kinetics. 

In the present model framework, which is adopted from the original envelope model, the evolving temperature field is obtained from a transient heat conduction equation that incorporates the moving heat source and the latent heat released inside the mushy zone, following the treatment in \cite{uddagiri_phase-field_2023}. The envelop surface is not treated as an iso-surface but rather in terms of diffuse phase field interface and velocity of the envelop surface is calculated by drawing an analogy between equation describing evolution of Phase-field and Gibbs-Thomson relation. The envelope motion is driven by the local undercooling, which contains thermal and solutal components. A full solution of solute diffusion at the scale of the dendrite tip would require spatial and temporal resolutions far below those feasible for melt pool simulations. Instead, a reduced microstructural description is employed, based on microscopic solvability theory for dendritic growth in dilute binary alloys. This model provides the dendrite-tip velocity as a function of total undercooling, which is subsequently converted into a normal envelope velocity for the phase-field representation.

The reason for opting for phase field based front propagation is that phase field equation has one significant advantage compared to models that solve the free boundary problem of solidification using adaptive finite elements, nowadays referred to as `front-tracking' models. The advantage is that the vector equation of a solidification front moving with envelope velocity $\vec v_{env}$ is replaced by a scalar `rate' equation using the identity 
\begin{equation} \label{advection}
\dot\phi(x,t) = \nabla \phi(x,t) \vec v_{env} \approx \frac{\pi}{\eta}\sqrt{\phi(1-\phi)} v_{env} . 
\end{equation} 
When $ \phi(x,t)$, which has a smeared-out or diffused step profile between the phases, grows from $0 \rightarrow 1$, the front moves by a distance $\eta$, which is the width of the diffuse interface in the direction normal to the interface 

The explicit form of the phase field equation with double obstacle potential is written as, 

\begin{equation}
   \label{pf2}
\dot\phi = M^{\phi} \sigma^* \left[\nabla^2\phi+\frac{\pi^2}{\eta^2}\left(\phi-\frac12\right)\right] + \frac{\pi}{\eta}\sqrt{\phi(1-\phi)} v_{env},  
\end{equation}
where $M^{\phi}$ is the interface mobility, $\sigma^{*}$ is the interface stiffness. Here we must pause for a moment. In standard phase-field models the first term proportional to interface stiffness and mobility represents capillarity, the velocity of the front, here $v_{env}$, is determined by the driving force for transformation and again the interface mobility. All these terms are ill defined in an envelope model of grain growth. To remove the spurious effect of the capillarity term, but keeping the interface stabilization active, in this work the product of interface stiffness and mobility has been chosen with a low value not to influence the results significantly (see Table \ref{tab:data}). For a rigorous treatment see \cite{Uddagiri2024}. The model for the envelope velocity is described below (\ref{v_env}).

The temperature field on the mesoscopic scale is obtained by solving a transient heat conduction equation with heat flux from the moving beam and latent heat release in the mushy zone. The governing equation is:

\begin{equation} \label{eq_T}
\rho C_{p}^{*} \dot T= \nabla (\lambda\nabla T) + \dot Q_{local} ,
\end{equation}

where $\rho$ is mass density, $C_{p}^{*}$ is effective heat capacity taking the Latent heat into consideration $C_p^* = C_p + L\frac{\partial f_s}{\partial T}$, $\lambda$ is effective thermal conductivity, L is solidification latent heat, $\dot f_{s}$ is local solid fraction and $Q_{local}$ is the combined heat flux resulting from melting and cooling. In order to account for release of latent heat in mushy zone heat capacity term is modified by assuming solid fraction to vary linearly between solidus and liquidus region. This gives $C_{p}^{*}$ is effective heat capacity taking the Latent heat into consideration $C_p^* = C_p + \frac{L}{\Delta T_0}$, where $\Delta T_0$ is the solidification window.

From the resulting temperature distribution, the total undercooling at each point is
\begin{equation}
\Delta T_{total} = T_{m} - T^{}.
\end{equation}

The microscopic solvability theory gives the dendrite-tip velocity as
\begin{equation}
v_{tip} = f(\Delta T_{total}, d_{0}, \sigma^{*}) = a \Delta T^n,
\end{equation}
where $a$ is the material constant which is a function of $d_{0}$, the capillary length and interface energy, $\sigma ^*$ . The envelope velocity is related to the tip velocity through the crystallographic growth direction and local interface orientation,
\begin{equation}
\label{v_env}
\vec v_{env} = \boldsymbol{\vec n} v_{tip} \cos\theta,
\end{equation}
where $\theta$ is the angle between the tip direction and the envelope normal.

\section{Simulation setup and Model parameters}

The model is implemented in open source Open-Phase library, which already has dedicated modules for phase field, temperature and crystal orientations. Simulation box is initialised with a substrate containing randomly oriented grains using voronoi tessellation. The heat source is modelled as Gaussian heat source. In additive manufacturing, when laser beam is focused on to the powder surface, depending on the beam power, energy is transmitted till certain powder bed depth. The incident beam undergoes several internal reflections and liquid is formed at the certain depth from surface {\cite{dorussen_efficient_2023}}. No flux boundary condition is applied at top surface and rest three surfaces are treated free boundaries to obtain constant temperature gradient across the surface. Here, the heat conductivity is treated as a free parameter to control shape and size of melt pool. Following this, liquid is nucleated approx.30 to 40 $\mu m$ below from top surface of the substrate. Unless otherwise specified or mentioned, all the simulations were performed with the data given in the table \ref{tab:data}:

\begin{table}[H]
    \centering
    \caption{Input parameters of the phase-field.}
    
    \begin{tabular}{cccc}
        \hline
        Parameter& Symbol& Value& Unit \\
         \hline
         Grid spacing& dx& 1& $\mu m$ \\
         Time step& dt& 0.1& $\mu s$ \\
         Interface width& $\eta$& 5.0& $\mu m$ \\
         Interface energy anisotropy solid-liquid & $\epsilon$& $0.35$& - \\
         Interface stiffness liquid-solid & $\sigma_{01}$& 0.24&$J.m^{-2}$\\
         Interface mobility between liquid-solid& $\mu_{01}$& $8.00\times10^{-7}$& $m^{3}s^{-1}N^{-1}$ \\
         Radii of Gaussian heat source&-& $11\times0\times17$&$\mu m$\\
         Velocity of Gaussian heat source&-& $9.0e^{-2}$ along x-direction&$m s^{-1}$\\
         Heat extraction rate due to irradiation&-&$6.3\times10^{2}$&$W cm^{-2}$\\
         Mole fraction of Ni &Ni&$0.795$&mole fraction\\
         Mole fraction of Al &Al&$0.205$&mole fraction\\
         Equilibrium Liquidus Temperature &-&$1675,36$&K\\
         Equilibrium Solidus Temperature &-&$1651,37$&K\\
         Melting Temperature of pure Ni&-&$1728,26$&K\\
         Liquidus slope&-&$-258.048$&-\\
         Solidus slope&-&$-375.073$&-\\
         
         \hline 
    \end{tabular}
     
    \label{tab:data}
\end{table}

\section{Results}

\subsection{2D grain growth simulations}

\begin{figure}[H]
    \centering
    \includegraphics[width = \textwidth,scale=1.8]{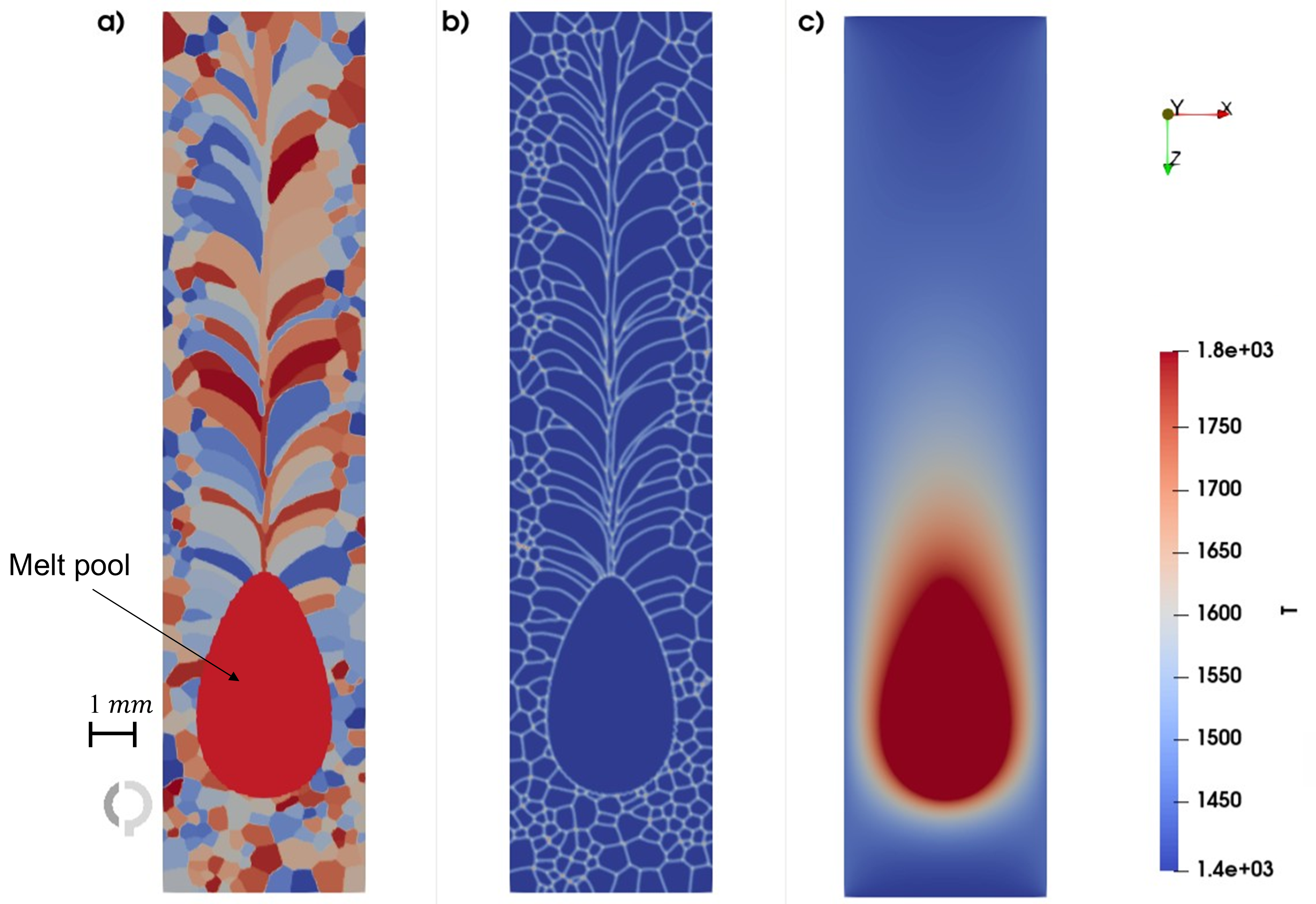}
    \caption{2D simulated grain structure due to single track melting (a) Simulated grain structure along with the melt pool geometry. (b) Grain interfaces (c) Temperature distribution towards the end of melting }
    \label{fig:single_track}
\end{figure}

We present here the 2D grain growth simulations using the proposed model. The simulations were conducted in a domain of 3 mm × 8.5 mm while the material properties remain the same as given in the table \ref{tab:data}. 
Figure \ref{fig:single_track} illustrates the temperature and corresponding grain evolution  due to melting and re-solidification under additive manufacturing process conditions. While Fig. \ref{fig:single_track} (a) shows the simulated grain structure along with the melt pool geometry, Fig. \ref{fig:single_track} (b) illustrates the grain structure and Fig. \ref{fig:single_track} (c) shows the temperature distribution towards the end of melting. From this figure, one can see that, a distribution of crystallographic orientations is initialized at the melt pool boundary, and grains advance epitaxially into the undercooled liquid. The resulting microstructure exhibits strong columnar growth aligned with the thermal gradient, along with competitive growth of the grains that result in the final textured microstructure.

\subsection{3D grain growth simulations}

\begin{figure}[H]
    \centering
    \includegraphics[width = \textwidth,scale=1.8]{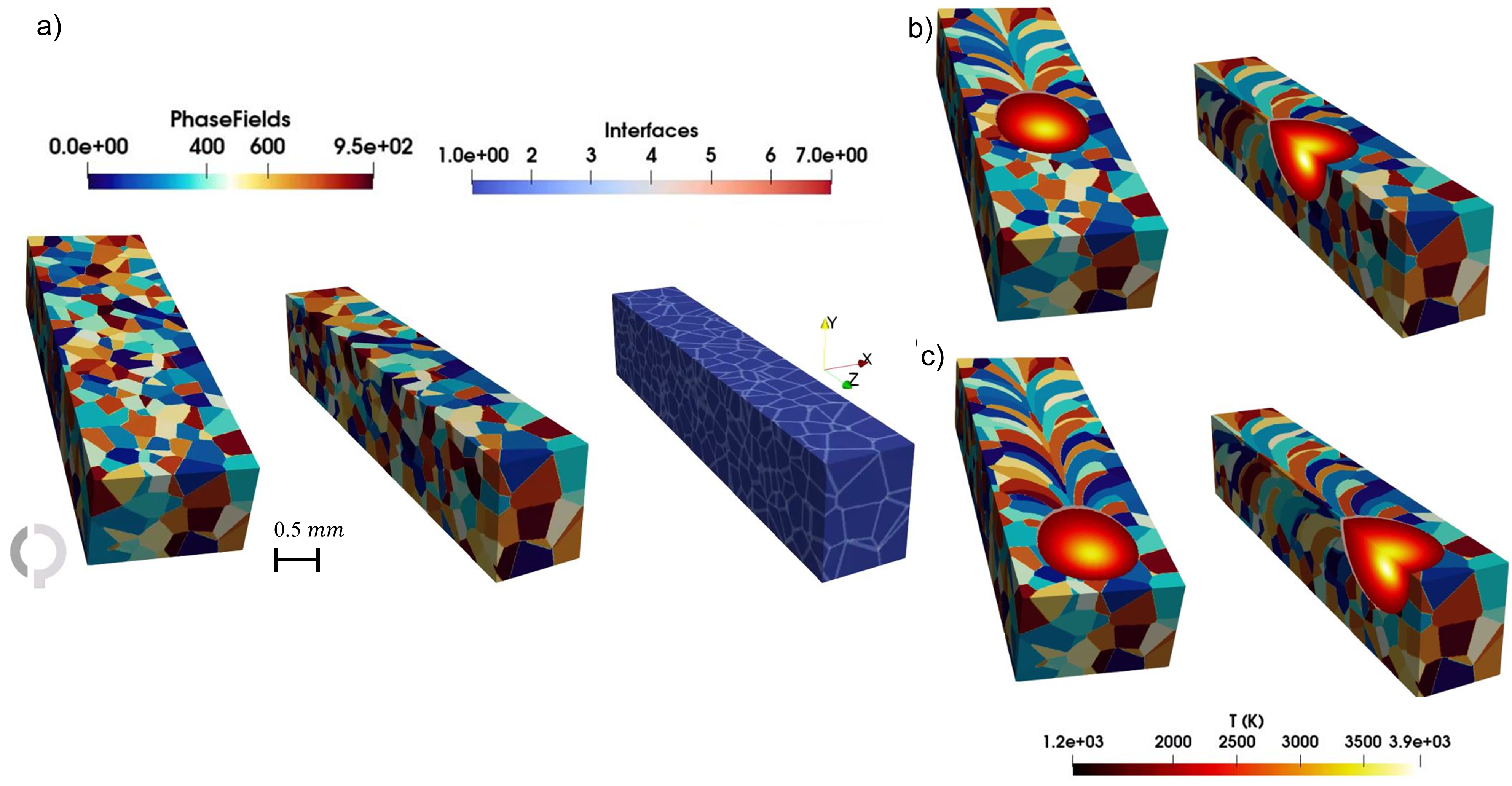}
    \caption{ 3D simulated grain structure due to single track melting (a) Initial microstructure (b) Intermediate stage of melting (c) Final stage of melting}
    \label{fig:3D_single_track}
\end{figure}

The three-dimensional simulations were performed in a domain of 1.5 mm × 1.2 mm × 5 mm, enabling full spatial reconstruction of melt-pool geometry, solid–liquid interface evolution, and competitive grain growth. Fig.  \ref{fig:3D_single_track}(a–c) shows the microstructure evolution during single track melting. While, Fig. \ref{fig:3D_single_track}(a) shows the initial grain structure created with voronoi grain generator algorithm, Fig.\ref{fig:3D_single_track}(b–c) show intermittent and final stages of melting. As in the 2D case, grain growth is governed by the local undercooling, yet the 3D simulation exhibits additional complexity due to out-of-plane competition of grain growth and curvature effects that cannot be captured in 2D simulations. The advancing envelopes organize into columnar structures aligned with the thermal gradient, while lateral growth is suppressed through competitive growth and impingement. The melt pool retains a characteristic elliptical morphology, and the network of interfaces reveals the expected refinement of grains near the melt pool boundary and coarsening around the solidification front.

\subsection{Parametric study: Influence of Material and Process Parameters on the Undercooled region}


In AM, crystallographic texture and the CET (Columnar to Equiaxed Transition) are directly linked to the local solidification conditions for heterogeneous or spontaneous nucleation ahead of the advancing solid–liquid interface.
For AM components, which are designed to have equiaxed morphology, experimental studies have frequently attempted to promote nucleation through inoculants, ultrasonic agitation, modified scan strategies, or controlled thermal management. Since the extent of constitutional undercooling governs the probability of new grain formation, quantifying the size and intensity of the undercooled zone is central for understanding microstructure selection and the onset of columnar-to-equiaxed transition (CET).

In this section, the developed mesoscopic model is utilized to study how the material parameters (specifically the kinetic coefficient $a$) and process parameters (captured through the resulting temperature gradient) influence the formation of the undercooled region ahead of the solidification front. Both factors directly influence the tip velocity predicted by the microscopic-solvability growth law and the resulting texture evolution.

In AM, thermal gradient (G) can be manipulated through beam power, scan strategy, layer preheating, or substrate temperature. Here, we vary the substrate temperature $T_{sub}$ while adjusting beam power to maintain comparable melt-pool dimensions. This approach reflects practical scenarios such as multi-beam preheating, where the substrate is preheated  prior to melting.

\begin{figure}[H]
    \centering
    \includegraphics[width = \textwidth,scale=1.8]{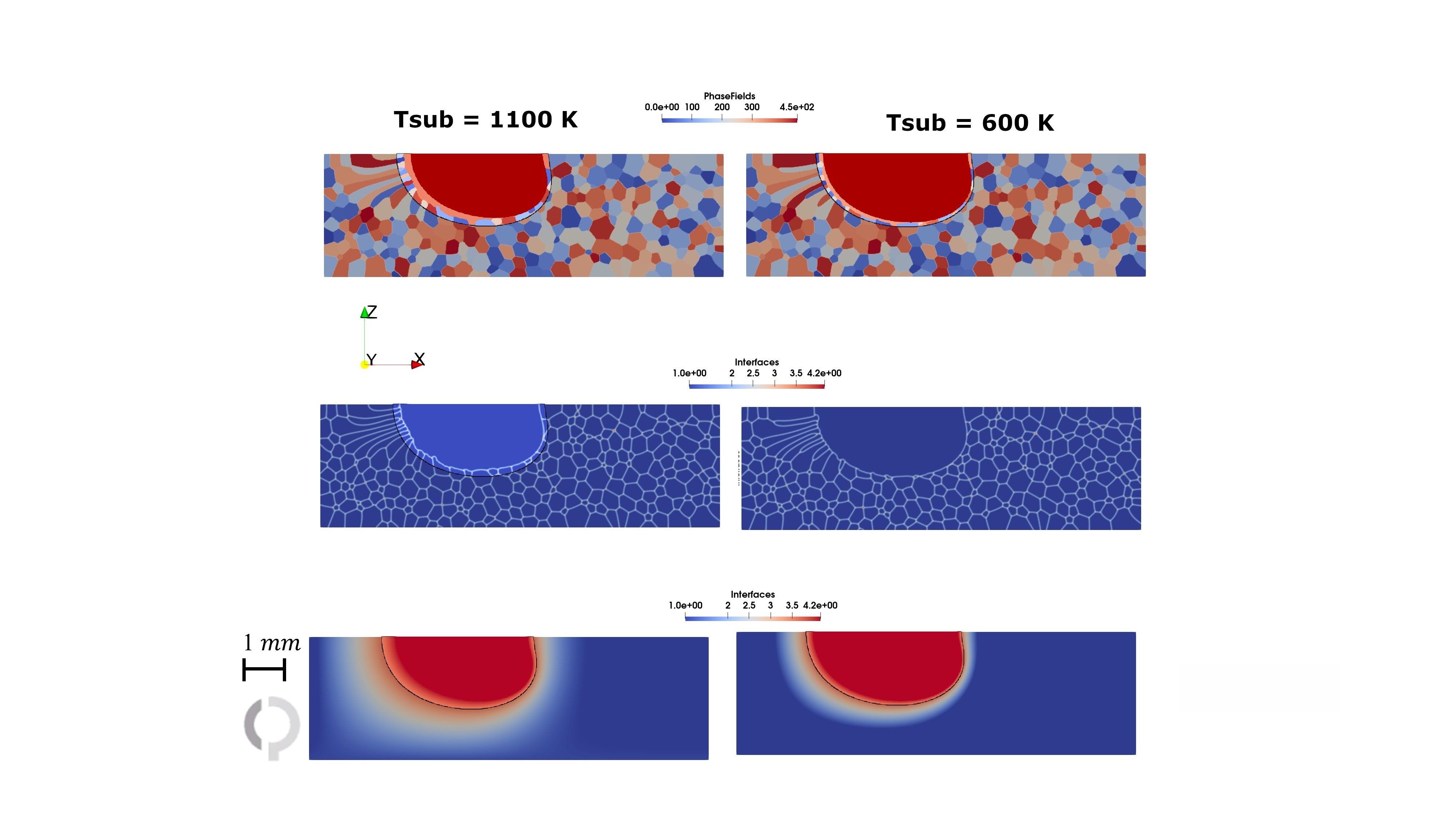}
    \caption{ Difference in undercooled region as a result of substrate temperature a) 1100 K and b) 600 K for the material coefficient a= 2500}
    \label{fig:grad_2500}
\end{figure}

\begin{figure}[H]
    \centering
   \includegraphics[width = \textwidth,scale=1.8]{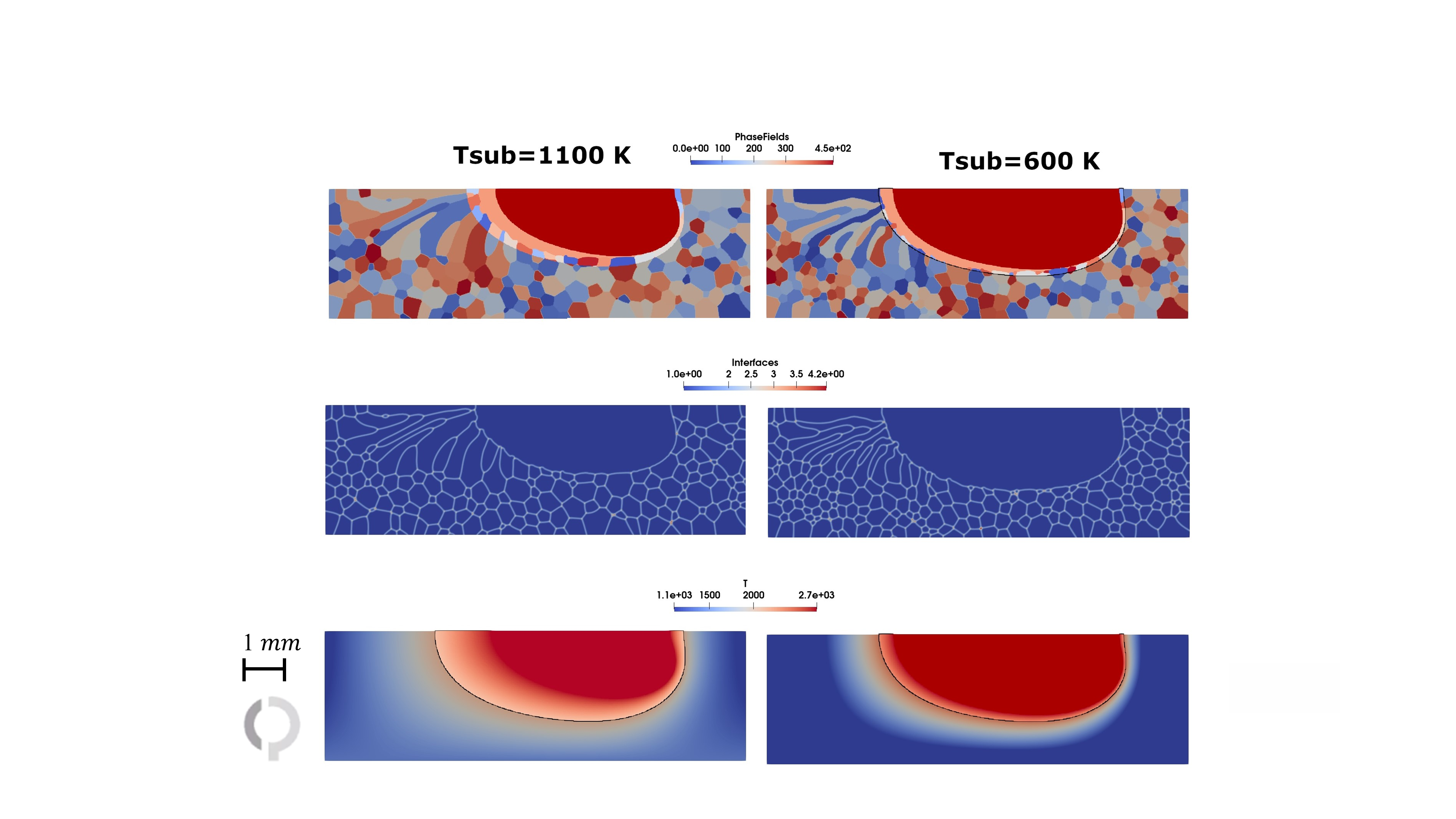}
    \caption{ Difference in undercooled region as a result of substrate temperature a) 1100 K and b) 600 K for the material coefficient a= 4500}
    \label{fig:grad_4500}
\end{figure}

Figures~\ref{fig:grad_2500} \& \ref{fig:grad_4500} illustrate the resulting undercooled regions for different substrate temperatures at a fixed material kinetic parameter $a$. For both the values of $a$, lowering the substrate temperature increases the thermal gradient and produces a narrower yet more intense undercooled region immediately ahead of the solidification front. This trend signifies that under steep thermal gradients and high interface velocities, solute rejection becomes increasingly suppressed due to reduced time for atomic diffusion. Consequently, the interface approaches the regime of absolute stability, where morphological perturbations are damped and dendrite envelopes advance more uniformly. These observations highlight the sensitivity of nucleation potential and CET behavior to both material kinetics and process-induced thermal conditions.

\subsection{Evolution of grain growth during multi-layer AM-Buildup}

Having established the ability of the model to simulate the grain growth in 2D and 3D for a single scan melting, we extend the model to simulate microstructure evolution during multilayer deposition. This enables modelling the resulting texture development that operate over many thermal cycles in PBF-AM.

\begin{figure}[H]
    \centering
    \includegraphics[width = \textwidth,scale=1.8]{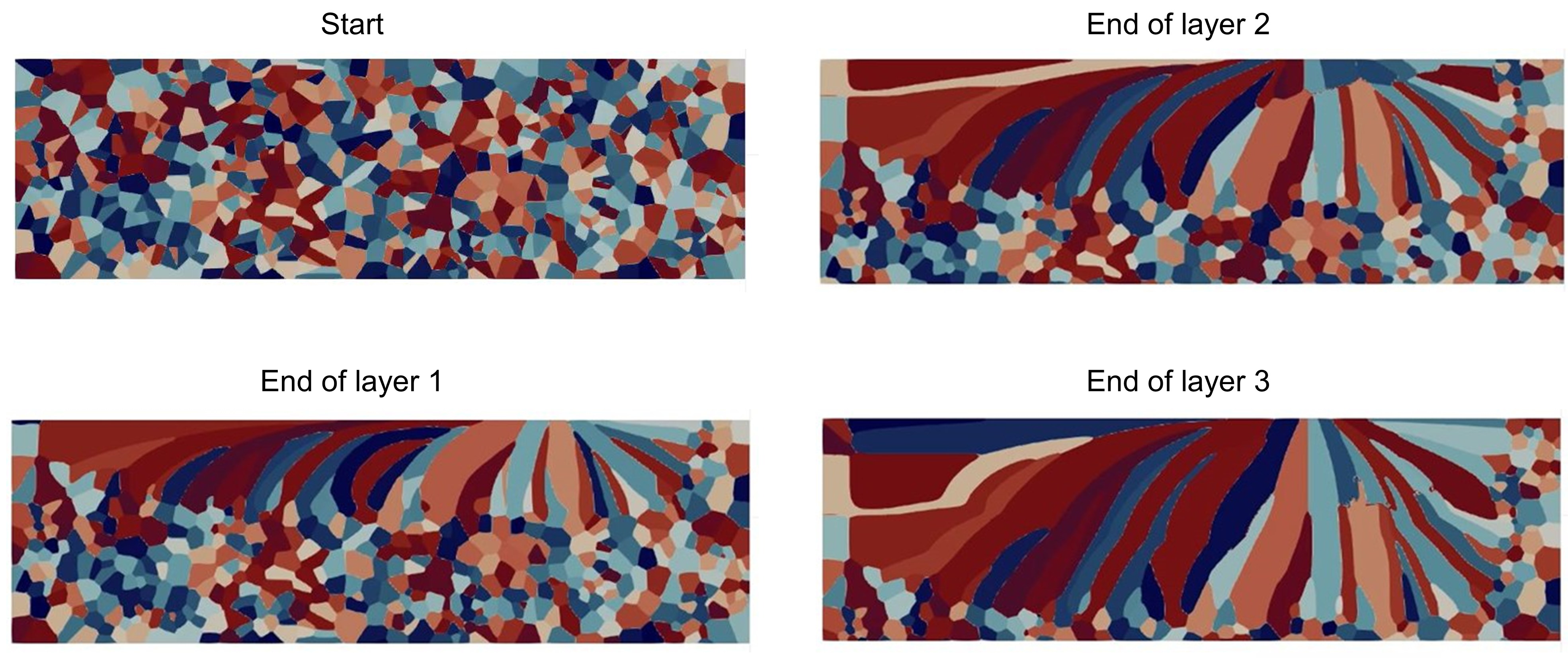}
    \caption{ Texture evolution during multi-track processing}
    \label{fig:AM_multitrack}
\end{figure}

Figure~\ref{fig:AM_multitrack} illustrates the evolution of grain morphology and crystallographic texture during three consecutive layer depositions. The initial microstructure is generated through a voronoi based distributed poly crystalline grain generator algorithm, resulting in an equiaxed polycrystalline grains with a broad orientation distribution. After the first layer is deposited, the large directional thermal gradients promote competitive grain growth. Therefore, grains with favorable alignment to the solidification direction advance preferentially, while misoriented grains are overgrown. By the second layer, epitaxial growth becomes clearly established, and columnar grains begin to extend across the melt pool, progressively overgrowing the equiaxed grain distribution in the underlying layer. The repeated thermal cycles associated with remelting, further intensify this selection, as only a small subset of crystallographic orientations continue to grow upward. During the deposition of third layer, this trend becomes even more pronounced. A limited number of dominant grain orientations survive and grow vertically, producing long columnar grains that span multiple layers. This behavior reflects the cumulative influence of thermal gradients and directional solidification conditions, whereby each new layer reinforces the existing texture. As a result, the initial random grain structure transitions toward a strong columnar texture, consistent with experimentally observed epitaxial grain growth in PBF-AM \cite{koepf20183d, gunasegaram_modelling_2021}.

Overall, the multilayer simulations demonstrate that the proposed phase-field approach captures the essential selection mechanisms governing microstructural inheritance and texture development in additive manufacturing.
\section{Summary \& Conclusions}

In this work, a mesoscopic grain-envelope model applying a phase-field front-propagation method was developed to simulate grain growth and texture evolution under additive manufacturing conditions. The model integrates transient heat conduction with moving heat sources and latent-heat release, and employs a microscopic-solvability-based kinetic law to relate local undercooling to envelope velocity. This approach provides a computationally efficient alternative to fully dendritic phase-field models while maintaining key solidification physics relevant to AM.

The model was first employed to perform both 2D and 3D simulations of microstucture evolution during single scan melting and re-solidification. These results demonstrated the capability of the model to capture epitaxial grain growth, competitive orientation selection, and the emergence of columnar textures under PBF-AM process conditions. Further, the parametric studies on substrate temperature and kinetic material coefficients revealed their strong influence on the extent of the undercooled region ahead of the solidification front, thereby controlling nucleation probability and  resulting columnar-to-equiaxed transition. The simulations results have shown that higher temperature gradients and larger growth velocities results in narrow undercooled zone, driving the grain growth toward the regime of absolute stability.

Finally, extension of the model to multi-layer build-up reproduced the evolution of characteristic texture across layers, competitive grain growth, and the formation of long, continuous columnar grains, which are well documented in AM processes. To conclude, these results demonstrate that the proposed mesoscopic envelope–phase-field methodology offers an effective mesoscale modelling alternative for predicting grain morphology and texture evolution in additive manufacturing, suitable for integration with process design and optimization frameworks.

\vspace{6pt}

\section*{conflicts of interest}
The authors declare no conflicts  of interest.

\bibliography{references}
\bibliographystyle{unsrt}

\end{document}